\documentclass[12pt,prd,tightenlines,nofootinbib]{revtex4}%[12pt,prd,showpacs,tightenlines,nofootinbib]{revtex4}
\usepackage{bm}
\usepackage{graphics}
\usepackage{epsfig}
\begin{document}

\title{
The possible candidates of tetraquark : $Z_b(10610)$ and $Z_b(10650)$}
\author{Tao Guo}
\email{guotao19870@126.com}
\author{Lu Cao}
\email{clcl@swu.edu.cn}
\author{Ming-Zhen Zhou}
\email{zhoumz@swu.edu.cn}
\author{Hong Chen\footnote{Corresponding Author}}
\email{chenh@swu.edu.cn}
\affiliation{School of Physical Science and Technology, Southwest
University, Chongqing 400715, China}

\begin{abstract}
Using the chromomagnetic interaction Hamiltonian with proper account for the $SU(3)$ flavor symmetry breaking,
we have performed a schematic study on the masses of $S-$wave heavy tetraquarks as $bq\bar{b}\bar{q}$ ($q$
denotes $u$, $d$, $s$ quark). It is found that the numeral results for $bu\bar{b}\bar{d}$ or $bd\bar{b}\bar{u}$
with $1^{+}$ quantum number are 10612 MeV and 10683 MeV respectively, which are well compatible with the recent
detected charged bottomonium-like $Z_b(10610)$ and $Z_b(10650)$. Theoretically, we also investigate the possible
tetraquark states of $1^{++}$ and $2^{+}$ due to the charge conjugation as the potential candidates for the updating
 experiments.
\end{abstract}

\maketitle

\section{Introduction}

In the recent discovery from the Belle Collaboration, two charged bottomonium-like resonances have been announced,
dubbed $Z_b(10610)$ and $Z_b(10650)$, in the $\pi^{\pm}\rm{\Upsilon}(nS)\, (n=1,2,3)$ and $\pi^{\pm}h_b(mP)\,(m=1,2)$
mass spectra \cite{Zb}. In detail, the weighted average masses and widths are $M[Z_b(10610)]=10608.4\pm2.0\rm{MeV}$,
$\rm{\Gamma}[Z_b(10610)]=15.6\pm2.5\rm{MeV}$ and $M[Z_b(10650)]=10653.2\pm1.5\rm{MeV}$,
$\rm{\Gamma}[Z_b(10650)]=14.4\pm3.2\rm{MeV}$ respectively.
Exclude the hypotheses of $J\leq2$ via the comparison of angular distributions, the quantum numbers for
both $Z_b(10610)$ and $Z_b(10650)$ favored $1^+$ \cite{Zb}. The fact that the non-zero electric charged
states cannot be explained as conventional quarkonium systems, has been encouraged great interest as exotic
states. Differ with the previous theoretical works \cite{Zb,Zhang,Bondar,Chen,Yang} exploring the two states as
 a meson-meson molecules as well as the mixture of molecules with bottomonium \cite{Bugg}, we provide a trivial
 research within the tetraquark framework regarding the two charged states as the possible candidates of tetraquark.

The possible existence of tetraquark states has recently received a renewed interest due to the new resonances in
the spectrum of states with open and hidden charm. The picture of tetraquark was foremost raised and was used to
describe scalar mesons below 1GeV in 1977 by Jaffe\cite{Jaffe1,Jaffe2,Jaffe3}. In the special models,
it is suggested that the $X(3872)$ can be a tetraquark state of $cq\bar{c}\bar{q}$, where $q$ denotes a light quark,
with $1^{++}$ quantum numbers\cite{Maiani1,Hogaasen,Buccella,Stancu1}. Besides $X(3940)$ is predicted as a possible
 tetraquark state of $2^{++}$\cite{Maiani1}, explicitly, $Y(4140)$ can be interpreted as a ground state of
 $cs\bar{c}\bar{s}$ \cite{Stancu2} as well as $Y(4260)$ particle can be the first orbital excitation of a
 diquark-antidiquark state $cs\bar{c}\bar{s}$ \cite{Maiani2}. Furthermore, another interesting the $Z(4430)$
 state, which has positive or negative charge, has been regarded as the first radial excitation of $cu\bar{c}\bar{d}$
 or $cu\bar{c}\bar{d}$ \cite{Maiani3,Rosner,Bracco} etc. If these states can be confirmed as tetraquarks, similar with
 the  charmonium-like $X(3872)$ assuming  $cq\bar{c}\bar{q}$ structure, tetraquark framework also holds with the
 possibility of $bq\bar{b}\bar{q}$ to the bottomonium-like states.

In this work, we perform a detailed study of the mass splitting of the $bq\bar{b}\bar{q}$ tetraquarks for
$J^{P}=1^+$ and $J^{P}=2^+$ quantum numbers in the $SU(3)\otimes SU(2)$ color-spin representations. In the
following part, we briefly review the simple quark model involved in the chromomagnetic interaction Hamiltonian
matrix. Then, the mass spectrum of $bq\bar{b}\bar{q}$ tetraquarks are calculated with account for flavor-symmetry
breaking. Some discussions are summarized in Sec.IV, as a brief conclusion.

\section{The tetraquark mass spectrum of type $bq\bar{b}\bar{q}$}

The chromomagnetic interaction arising from one-gluon exchange in MIT bag model is quite successful in
the description the mass splitting of meson and baryon spectrum \cite{Rujula}. The interaction Hamiltonian
acting on the colour and spin degrees of freedom reads \cite{Jaffe2,Rujula}:
\begin{equation}\label{colourmagnetic}
H_{cm}=-\sum\limits_{i>j}v_{ij}\tilde{\lambda}_{i}^{c}\cdot\tilde{\lambda}_{j}^{c}
\vec{\sigma}_{i}\cdot\vec{\sigma}_{j},
\end{equation}
where $\vec{\sigma}_i$ is the quark spin operator and $\tilde{\lambda}_{i}^{c}$ is the color operator for
the $i$-th quark. The values of the coefficients $v_{ij}$ depend on the quark masses and properties of the
special wave function of the multiquark system. To be precise, if $i$ or $j$ indicates an antiquark, the
following replacement should be understood:
\begin{equation}
\tilde{\lambda}_{i}\rightarrow -\tilde{\lambda}_{i}^{\ast},\quad
\vec{\sigma}_{i}\rightarrow -\vec{\sigma}_{i}^{\ast}.
\end{equation}
Accordingly, the mass of tetraquark is given by the expectation value of the colourmagnetic model
Hamiltonian \cite{Hogaasen}:
 \begin{equation}\label{Hamiltonian}
 H=\sum\limits_im_i+H_{cm},
 \end{equation}
where $m_i$ is effective mass of the $i$-th constituent quark. Subsequently, the $m_i$ in Eq.(\ref{Hamiltonian})
and $v_{ij}$ in Eq.(\ref{colourmagnetic}) as parameters can be extracted from heavy mesons in the following
calculation.

Here we simply use chromomagnetic interaction to fit heavy mesons experimental masses from which the
constituent effective mass $m_i$ and the coefficient $v_{ij}$ can be obtained for the $bq\bar{b}\bar{q}$ tetraquark
system. Using the corresponding experimental values of $J/\Psi$, $\eta_c$, $\rm{\Upsilon}$, $\eta_b$,
$B^\ast$, $B^{\pm}$, $B^\ast_s$, $B_s^0$, and $B_c^{\pm}$ masses \cite{Nakamura}. We use Eq.(\ref{Hamiltonian}) to
 extract these parameters:
\begin{equation}
\left.
\begin{array}{ll}
m_{J/\Psi}=2m_c+\frac{16}{3}v_{c\bar{c}},&\quad m_{\eta_c}=2m_c-16v_{c\bar{c}},\\
m_\Upsilon=2m_b+\frac{16}{3}v_{b\bar{b}},&\quad m_{\eta_b}=2m_b-16v_{b\bar{b}},\\
m_{B^{\ast}}=m_u+m_b+\frac{16}{3}v_{b\bar{u}},&\quad m_{B^\pm}=m_u+m_b-16v_{b\bar{u}},\\
m_{B_{s}^{\ast}}=m_s+m_b+\frac{16}{3}v_{b\bar{s}},&\quad m_{B_{s}^0}=m_s+m_b-16v_{b\bar{s}},\\
m_{B_c^\pm}=m_c+m_b-16v_{b\bar{c}}.&
\end{array}
\right.
\end{equation}
Therefore, we have
\begin{equation}\label{parameters}
\left.
\begin{array}{llll}
m_u=592.14{\rm{MeV}},&\quad m_s=681.65{\rm{MeV}},&\quad m_c=1533.88{\rm{MeV}},&\quad m_b=4721.47{\rm{MeV}},\\
v_{c\bar{c}}=5.47{\rm{MeV}},&\quad v_{b\bar{u}}=2.15{\rm{MeV}},&\quad v_{b\bar{s}}=2.30{\rm{MeV}},&\quad v_{b
\bar{c}}=4.06{\rm{MeV}},\\
v_{b\bar{b}}=3.25{\rm{MeV}},&&&
\end{array}
\right.
\end{equation}
where $v_{c\bar{c}}$, $v_{b\bar{u}}$, $v_{b\bar{s}}$, $v_{b\bar{c}}$, and $v_{b\bar{b}}$ are the coefficient
in Eq.(\ref{colourmagnetic}) for $c\bar{c}$, $b\bar{u}$, $b\bar{s}$, $b\bar{c}$, and $b\bar{b}$ meson system
respectively. On the other hand, we can easily obtain $v_{q\bar{q}}=29.8{\rm{MeV}}$ from fitting the experimental
values of light mesons.

Below we define the total color-spin wave function of tetraquark and give the colormagnetic Hamiltonian matrix.
There are a set of $J^{P}=0^+$, $1^+$ and $2^+$ quantum numbers for the possible ground state tetraquark.
In the case of color-singlet, all conceivable color-spin states have been constructed for given these quantum
numbers \cite{Stancu2}. Specially, we use the same notation which has been introduced in Ref. \cite{Hogaasen}
for the basis vectors and only write $J^P=1^+$ and $2^+$ states in the work. For $J^P=1^+$, the basis can be built
with $(1,3)$ and $(2,4)$ subsystems:
\begin{equation}\label{1324basis1}
\begin{array}{cc}
\alpha_1=[(q_1\bar{q}_3)^1_0\otimes(q_2\bar{q}_4)^1_1]^{1}_{1}, &\quad \alpha_2=[(q_1\bar{q}_3)^1_1\otimes(q_2\bar{q}_4)^1_0]^{1}_{1},\\
\alpha_3=[(q_1\bar{q}_3)^1_1\otimes(q_2\bar{q}_4)^1_1]^{1}_{1}, &\quad \alpha_4=[(q_1\bar{q}_3)^8_0\otimes(q_2\bar{q}_4)^8_1]^{1}_{1},\\
\alpha_5=[(q_1\bar{q}_3)^8_1\otimes(q_2\bar{q}_4)^8_0]^{1}_{1}, &\quad \alpha_6=[(q_1\bar{q}_3)^8_1\otimes(q_2\bar{q}_4)^8_1]^{1}_{1},
\end{array}
\end{equation}
where the superscript and subscript indicate a well defined color and spin, respectively. For $bq\overline{bq}$
($q$ and $\bar{q}$ denote quark and anti-quark of the same flavor) states, where $\alpha_3$ and $\alpha_6$ have
charge conjugation $C=+1$ and $\alpha_1$, $\alpha_2$, $\alpha_4$, $\alpha_5$ have conjugation $C=-1$ as explained
in Ref. \cite{Stancu1,Stancu2}. Similarly, the state of $J^P=1^+$ can be also rewritten in the $(1,4)$ and $(2,3)$
basis, not given detailed form here, which can be easily gained from Ref. \cite{Hogaasen}.
The chromomagnetic interaction Hamiltonian $H_{cm}$ acting on this basis vectors (\ref{1324basis1}) can be written
the following blocks matrix
\begin{equation}
H_{cm}=-\left(
\begin{array}{ccc}
A_{11} & A_{12}\\
A_{21} & A_{22}
\end{array}
\right),
\end{equation}
with $3\times3$ submatrices
\begin{equation}
A_{11}=\left(
\begin{array}{ccc}
\frac{16}{3}(3v_{13}-v_{24})& 0 & 0\\
0 & -\frac{16}{3}(v_{13}-3v_{24}) & 0\\
0 & 0 & -\frac{16}{3}(v_{13}+3v_{24})
\end{array}
\right),
\end{equation}

\begin{equation}
A_{12}= A_{21}^{\dagger}=\left(
\begin{array}{ccc}
0 & \left.
\begin{array}{c}
\frac{4\sqrt{2}}{3}(v_{12}+v_{34}\\[-0.1em]
+v_{14}+v_{23})
\end{array}
\right.& \left.
\begin{array}{c}
-\frac{8}{3}(v_{12}-v_{34}\\[-0.1em]
-v_{14}+v_{23})
\end{array}
\right.\\[0.9em]
\left.
\begin{array}{c}
\frac{4\sqrt{2}}{3}(v_{12}+v_{34}\\[-0.1em]
+v_{14}+v_{23})
\end{array}
\right. & 0 & \left.
\begin{array}{c}
-\frac{8}{3}(v_{12}-v_{34}\\[-0.1em]
+v_{14}-v_{23})
\end{array}
\right.\\[0.9em]
\left.
\begin{array}{c}
-\frac{8}{3}(v_{12}-v_{34}\\[-0.1em]
-v_{14}+v_{23})
\end{array}
\right. & \left.
\begin{array}{c}
-\frac{8}{3}(v_{12}-v_{34}\\[-0.1em]
+v_{14}-v_{23})
\end{array}
\right. & \left.
\begin{array}{c}
-\frac{4\sqrt{2}}{3}(v_{12}+v_{34}\\[-0.1em]
-v_{14}-v_{23})
\end{array}
\right.
\end{array}
\right),
\end{equation}

\begin{equation}
A_{22}= \left(
\begin{array}{ccc}
-2v_{13}+\frac{2}{3}v_{24} &
\left.
\begin{array}{c}
-\frac{4}{3}(v_{12}+v_{34})\\[-0.1em]
+\frac{14}{3}(v_{14}+v_{23})
\end{array}
\right.&
\left.
\begin{array}{c}
\frac{4\sqrt{2}}{3}(v_{12}-v_{34})\\[-0.1em]
+\frac{14\sqrt{2}}{3}(v_{14}-v_{23})
\end{array}
\right.\\[0.9em]
\left.
\begin{array}{c}
-\frac{4}{3}(v_{12}+v_{34})\\[-0.1em]
+\frac{14}{3}(v_{14}+v_{23})
\end{array}
\right. &
\frac{2}{3}v_{13}-2v_{24} &
\left.
\begin{array}{c}
-\frac{4\sqrt{2}}{3}(v_{12}-v_{34})\\[-0.1em]
+\frac{14\sqrt{2}}{3}(v_{14}-v_{23})
\end{array}
\right.\\[0.9em]
\left.
\begin{array}{c}
\frac{4\sqrt{2}}{3}(v_{12}-v_{34})\\[-0.1em]
+\frac{14\sqrt{2}}{3}(v_{14}-v_{23})
\end{array}
\right. &
\left.
\begin{array}{c}
-\frac{4\sqrt{2}}{3}(v_{12}-v_{34})\\[-0.1em]
+\frac{14\sqrt{2}}{3}(v_{14}-v_{23})
\end{array}
\right. &
\left.
\begin{array}{c}
\frac{2}{3}(v_{13}+v_{24})+\frac{4}{3}(v_{12}+v_{34})\\
+\frac{14}{3}(v_{14}+v_{23})
\end{array}
\right.
\end{array}
\right).
\end{equation}

For $J^P=2^+$ the situation is more simple. There are two linearly independent basis vectors:
\begin{equation}\label{1324basis2}
\left.
\begin{array}{cc}
\beta_1=[(q_1\bar{q}_3)^1_1\otimes(q_2\bar{q}_4)^1_1]^{1}_{2}, &\quad
\beta_2=[(q_1\bar{q}_3)^8_1\otimes(q_2\bar{q}_4)^8_1]^{1}_{2}.\\
\end{array}
\right.
\end{equation}
The corresponding $H_{cm}$  acting on this basis (\ref{1324basis2}) can be written the $2\times2$ matrix
\begin{equation}
H_{cm}=-\left(
\begin{array}{ccc}
-\frac{16}{3}(v_{13}+v_{24})
&\left.
\begin{array}{c}
-\frac{4}{3}\sqrt{2}(v_{12}+v_{34}\\[-0.1em]
-v_{14}-v_{23})
\end{array}
\right.\\[0.9em]
\left.
\begin{array}{c}
-\frac{4}{3}\sqrt{2}(v_{12}+v_{34}\\[-0.1em]
-v_{14}-v_{23})
\end{array}
\right.
&\left.
\begin{array}{c}
\frac{2}{3}(v_{13}+v_{24})\\[-0.1em]
-\frac{4}{3}(v_{12}+v_{34})-\frac{14}{3}(v_{14}+v_{23})
\end{array}
\right.
\end{array}
\right).
\end{equation}
As such, the $(1,4)$ and $(2,3)$ basis are not given detailed form for the tetraquark states of $J^P=2^+$. If the
reader is interested in the question, please refer to the appendix part in Ref. \cite{Stancu2} where there is
accurate identification.

\section{Numerical results and discussion}

Here we respectively diagonalize the above $6\times6$ and $2\times2$ Hamiltonian matrix elements in the complete
tetraquark configuration space of type $bq\bar{b}\bar{q}$ with the $1^+$ and $2^+$ quantum numbers.
In the calculation of the matrix elements with account for the heavy quark limit, we approximately consider that
the parameters $v_{b\bar{q}}$ and $v_{bq}$ are equal because their difference is very small in the heavy quark limit.
Therefore, we can obtain the chromomagnetic interaction eigenvalues $(E)$ and the masses spectrum $(M)$ of tetraquarks
of type $bq\overline{bq}$ for $1^+$ quantum numbers which have been exhibited in Table \ref{1+table}. Furthermore,
we can observe there are masses of several states in the range $10550\sim10700$MeV. Concretely, we will discuss these
theoretical tetraquark states as the possible charged bottomonium-like resonances, both $Z_b(10610)$ and $Z_b(10650)$,
as follows.
\begin{table}[htbp]
\caption{Chromomagnetic interaction eigenvalues $(E)$, the masses $(M)$ and amplitudes of the basis
vectors (\ref{1324basis1})of $bq\overline{bq}$ with $1^{+}$ quantum numbers}
\label{1+table}
\vspace{0.2cm}
\centering
\begin{tabular}{cccccccccc}
 \hline \hline
$J^{P}$ & $E$(MeV) & $M$(MeV) &  $\alpha_1$ & $\alpha_2$ & $\alpha_3$ & $\alpha_4$ & $\alpha_5$ & $\alpha_6$ \\
 \hline
$1^{+}$ & -460.056& 10167.9 & -0.00003  & -0.99934& 0 & -0.03631 & -0.00101 & 0\\
$1^{+}$ & -15.670 & 10612.3 & -0.02616 & 0.03578 & 0 & -0.97924 &  -0.19780 & 0\\
$1^{+}$ & 55.348  & 10683.3 & 0.29403   & 0.00619 & 0 & -0.19657 & 0.93534 & 0\\
$1^{+}$ & 111.911 & 10739.9 & -0.95544  & 0.00096 & 0 & -0.03368 & 0.29327 & 0\\
$1^{+}$ & -47.833 & 10580.2 & 0         & 0       & 0 & 0        & 0 & 1 \\
$1^{+}$ & 176.267 & 10804.3 & 0         & 0       & 1 & 0        & 0 & 0 \\
 \hline \hline
\end{tabular}
\end{table}

For the $1^{+}$ states, as it can be seen from Table \ref{1+table}, there are three states
at 10580 MeV, 10612 MeV and 10683 MeV which are quite close to the experimental values of two
narrow charged structures $Z_b(10610)$ and $Z_b(10650)$. Among the three states, the lowest state
at 10580 MeV has an appropriate mass for the experimental range of $Z_b(10610)$, but the wave
function of this state only is $\alpha_6$, a pure octet-octet hidden color state, which obviously
cannot dissociate into a bottomonium state and a light pseudoscalar meson.
Thus the $Z_b(10610)$ is disfavored for the $1^{+}$ tetraquark state at 10580 MeV.
For both states of 10612 MeV and 10683 MeV, which are also commendably close to the experimental
range, their structures can be composed of $bu\bar{b}\bar{d}$ or $bd\bar{b}\bar{u}$.
As seen from Table \ref{1+table}, both states have a very small component on the state $\alpha_2$,
and therefore they have a narrow width  in $\rm{\Upsilon}\pi$ decay channel as well as they can be
taken as the best candidates for $Z_b(10610)$ and $Z_b(10650)$ from Belle Collaboration\cite{Zb}.

It is worthwhile to notice that the above three states are not the lowest-lying state for $bq\bar{b}\bar{q}$
system. The lowest state at 10167.9 MeV is strongly coupled to the channel consisting of a bottomonium state
and a light pseudoscalar meson because there is very big amplitude of the basis vector $\alpha_2$. Therefore,
this state is probably very broad and fall parts very easily. Thus it is not easy to observe it experimentally.

In the case of $q$ and $\bar{q}$ being the same flavor, the $1^{+}$ state can be separated $1^{++}$ and $1^{+-}$
states from the general consideration of charge conjugation symmetry. The lowest $1^{++}$ state at 10580 MeV is
entirely analogous to $X(3872)$ as $cq\bar{c}q$ tetraquark state. In Table I, $V_{bq}=V_{b\bar{q}}$ are assumed,
this states only is $\alpha_{6}$, a hidden color state, but if $\Delta=V_{bq}-V_{b\bar{q}}$ have a nonvanishing,
but small, value,
the state mass is nonsensitive to the change of $\Delta$ from Table \ref{1++table}. Meanwhile, this state shall
receive the small component coming from $\alpha_3$, as shown in Table \ref{1++table}, which implies
that this state can decay into $\rm{\Upsilon}+\rho$ or $\rm{\Upsilon}+\omega$ with the narrow width.

Such as we can predict there is a state of tetraquark of $2^{+}$ at 10631 MeV analogous to the $X(3940)$ \cite{Maiani1}.
As it can be seen from Table \ref{2+table}, the lowest state is a almost pure hidden color state in
the $b\bar{b}+q\bar{q}$ channel even if $v_{b\bar{q}}$ and $v_{bq}$ are not equal.
\begin{table}[htbp]
\caption{Chromomagnetic interaction eigenvalues $(E)$, the masses $(M)$ and amplitudes of the basis
vectors (\ref{1324basis1}) of the lowest $1^{++}$ ~$bq\overline{bq}$ state with the different $\Delta$.}
\label{1++table}
\vspace{0.2cm}
\centering
\begin{tabular}{cccccccccc}
 \hline \hline
$\Delta$(MeV) & $E$(MeV) & $M$(MeV) &  $\alpha_1$ & $\alpha_2$ & $\alpha_3$ & $\alpha_4$ & $\alpha_5$ & $\alpha_6$ \\
 \hline
-1.05 & -45.104 & 10582.9 & 0 & 0 & 0.01788 & 0 & 0 & 0.99984\\
1.05 & -50.702 & 10577.3 & 0 & 0 & -0.01744 & 0 & 0 & 0.99985\\
 \hline \hline
\end{tabular}
\end{table}
\begin{table}[htbp]
\caption{Chromomagnetic interaction eigenvalues $(E)$, the masses $(M)$ and amplitudes of the basis
 vectors (\ref{1324basis1}) of $bq\overline{bq}$ with $2^{+}$ quantum numbers}
\label{2+table}
\vspace{0.2cm}
\centering
\begin{tabular}{cccccccccc}
 \hline \hline
$J^{P}$ & $E$(MeV) & $M$(MeV) &  $\beta_1$ & $\beta_2$\\
 \hline
$2^{+}$   & 3.767 & 10631.0  & 0  & 1\\
$2^{+}$   & 176.267& 10803.5 & 1  & 0\\
 \hline \hline
\end{tabular}
\end{table}

\section{Summary}

In this work, the mass spectrum of tetraquark of type
$bq\overline{bq}$ with $1^+$ and $2^+$ quantum numbers have been
calculated by using the simple chromomagnetic interaction with
proper account for the $SU(3)$ flavor symmetry breaking. The
numerical result both 10612 MeV and 10683 MeV are well compatible
with the experimental values for describing the charged
bottomonium-like $Z_b(10610)$ and $Z_b(10650)$, which should have a
minimum tetraquark content $bu\bar{b}\bar{d}$ or $bd\bar{b}\bar{u}$,
respectively. At the same time, we also predict two possible
tetraquark states of $1^{++}$ and $2^{+}$ states, which have masses
of 10580 MeV and 10631 MeV, and expect to search them in future
experiments.

\section*{Acknowledgements}
This work is partly supported by the National Natural Science
Foundation of China under Grant Nos. 10575083, 11005087.

\end{document}